\documentclass[useAMS,usenatbib]{mn2e}
\usepackage{epsfig}
\usepackage{amssymb}
\usepackage{amsmath}
\usepackage{lscape}
\usepackage{longtable}
\usepackage[normal]{caption}
\usepackage{appendix}
\usepackage{bm}

\title[What powers Hyperluminous Infrared galaxies at $z$$\sim$1--2?]{What powers Hyperluminous infrared galaxies at $z$$\sim$1--2?}

\author[M.~Symeonidis and M. J. ~Page] 
{\parbox{\textwidth}{\raggedright
M.~Symeonidis$^{1}$\thanks{E-mail: \texttt{m.symeonidis@ucl.ac.uk}}
and M. J. ~Page$^{1}$ }\vspace{0.4cm}\\
\parbox{\textwidth}{\raggedright $^{1}$ Mullard Space Science
  Laboratory, University College London, Holmbury St. Mary, Dorking,
  Surrey RH5 6NT, UK}}

\begin{document}

\date{Accepted  Received; in original form}

\pagerange{\pageref{firstpage}--\pageref{lastpage}} \pubyear{2014}

\maketitle

\label{firstpage}

\begin{abstract}
We investigate what powers hyperluminous infrared galaxies (HyLIRGs; $L_{\rm IR, 8-1000\mu m}>$10$^{13}$\,L$_{\odot}$) at z$\sim$1-2, by examining the behaviour of the infrared AGN luminosity function in relation to the infrared galaxy luminosity function. The former corresponds to emission from AGN-heated dust only, whereas the latter includes emission from dust heated by stars and AGN. 
Our results show that the two luminosity functions are substantially different below 10$^{13}$\,L$_{\odot}$ but converge in the HyLIRG regime. We find that the fraction of AGN dominated sources increases with total infrared luminosity and at $L_{\rm IR}>10^{13.5}\, \rm L_{\odot}$ AGN can account for the entire infrared emission. We conclude that the 
bright end of the $1<z<2$ infrared galaxy luminosity function is shaped by AGN 
rather than star-forming galaxies. 
\end{abstract}


\section{Introduction}
\label{sec:introduction}

Hyperluminous infrared galaxies (HyLIRGs) are defined as galaxies with $8-1000$~$\mu$m infrared luminosities (hereafter $L_{\rm IR}$) exceeding $10^{13}$\,L$_{\odot}$. They were first discovered in the \textit{IRAS} all-sky survey (Rowan-Robinson et al. 1991\nocite{RR91}; Soifer et al. 1994\nocite{Soifer94}; Sanders $\&$ Mirabel 1996\nocite{SM96}) and the origin of their extreme luminosities immediately became a topic of much debate. HyLIRGs are rare; none have been found in the local Universe ($z<0.1$; e.g. Sanders et al. 2003\nocite{Sanders03}) and it is estimated that there are about 200 \textit{IRAS}-detected hyperluminous galaxies over the whole sky (Rowan-Robinson 2000\nocite{RR00}; Rowan-Robinson $\&$ Wang 2010\nocite{RRW10}). 

Follow-up studies of HyLIRGs found that their morphologies show signs of interactions or QSO-like appearance (e.g. Sanders et al. 1988\nocite{Sanders88}; Ivison et al. 1998\nocite{Ivison98}; Hines et al. 1999\nocite{Hines99}; Farrah et al. 2002a\nocite{Farrah02a}) and virtually all host AGN (e.g. Farrah et al. 2002b\nocite{Farrah02b}; Ruiz, Carrera $\&$ Panessa 2007\nocite{RCP07}; Ruiz et al. 2010\nocite{Ruiz10}; 2013\nocite{Ruiz13}), many of which are found to be buried (Hines et al. 1995\nocite{Hines95}; Granato et al. 1996\nocite{GDF96}). The observation that both AGN and starburst components are often necessary to fit their broadband spectral energy distributions (SEDs) has generated differing views on what the dominant energy source is in HyLIRGs: on one hand it has been suggested that emission longwards of 50$\mu$m is primarily attributed to star-formation, with star-formation rates that were predicted to be in excess of 1000\,M$_{\odot}$/yr (e.g. Rowan-Robinson et al. 2000\nocite{RR00}; Rowan-Robinson $\&$ Wang 2010\nocite{RRW10}; Verma et al. 2002\nocite{Verma02}), whereas on the other hand, the detection of buried, even Compton-thick, AGN in many HyLIRGs (e.g. Franceschini et al. 2000\nocite{Franceschini00}; Wilman et al. 2003\nocite{Wilman03}; Ruiz, Carrera $\&$ Panessa 2007\nocite{RCP07}; Stern et al. 2014\nocite{Stern14}) has given rise to the idea that all HyLIRGs could potentially host AGN and be powered by AGN (e.g. Hines et al. 1995\nocite{Hines95}; Granato et al. 1996\nocite{Granato96}; Yun $\&$ Scoville 1998\nocite{YS98}; Efstathiou et al. 2013\nocite{Efstathiou13}).

In this letter, we investigate what is the dominant power source in HyLIRGs, using the infrared galaxy luminosity function and the X-ray AGN luminosity function at $z\sim$1--2. This redshift range corresponds to the peak of the cosmic star-formation history and supermassive black hole accretion rate history (e.g. Marconi et al. 2004\nocite{Marconi04}; Madau $\&$ Dickinson 2014\nocite{MD14}). Moreover, the contribution to the cosmic energy budget from HyLIRGs is seen to undergo rapid growth from $z=0$ to $z=2$, amounting to at least a few percent of the total comoving infrared luminosity density at $z\sim2$ (e.g. B{\'e}thermin et al. 2011\nocite{Bethermin11}; Toba et al. 2015\nocite{Toba15}). 

This letter is laid out as follows: in section \ref{sec:method} we describe our method and our results are presented in section \ref{sec:results}. A discussion of our findings and conclusions are presented in sections \ref{sec:discussion} and \ref{sec:conclusions} respectively. Throughout, we adopt a concordance cosmology of H$_0$=70\,km\,s$^{-1}$Mpc$^{-1}$, $\Omega_{\rm M}$=1-$\Omega_{\rm \Lambda}$=0.3.

\begin{figure}
\epsfig{file=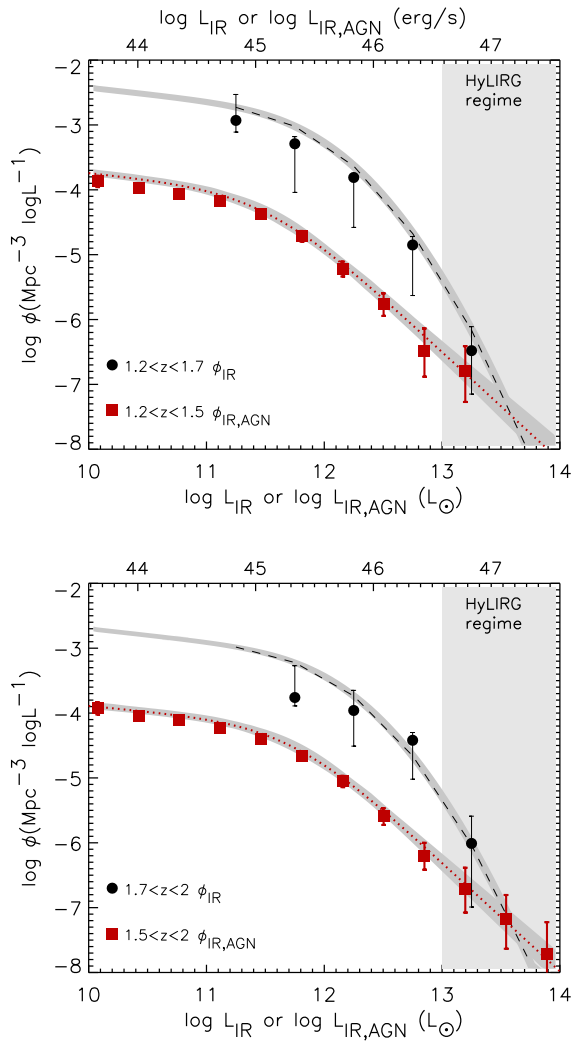,width=0.99\linewidth} 
\caption{The IR luminosity function from Gruppioni et al. (2013): black filled circles. The corresponding functional form and 1$\sigma$ uncertainty are shown by the black curve and shaded region. Shown with red squares is the IR AGN luminosity function, derived from the hard X-ray luminosity function in Aird et al. (2015). The red curve and shaded outline represents the functional form and 1$\sigma$ uncertainty. The 2 panels correspond to different redshift bins, with the shaded vertical band indicating the HyLIRG regime ($L_{\rm IR}>10^{13}$\,L$_{\odot}$). }
\label{fig:LFs}
\end{figure}

\begin{figure}
\epsfig{file=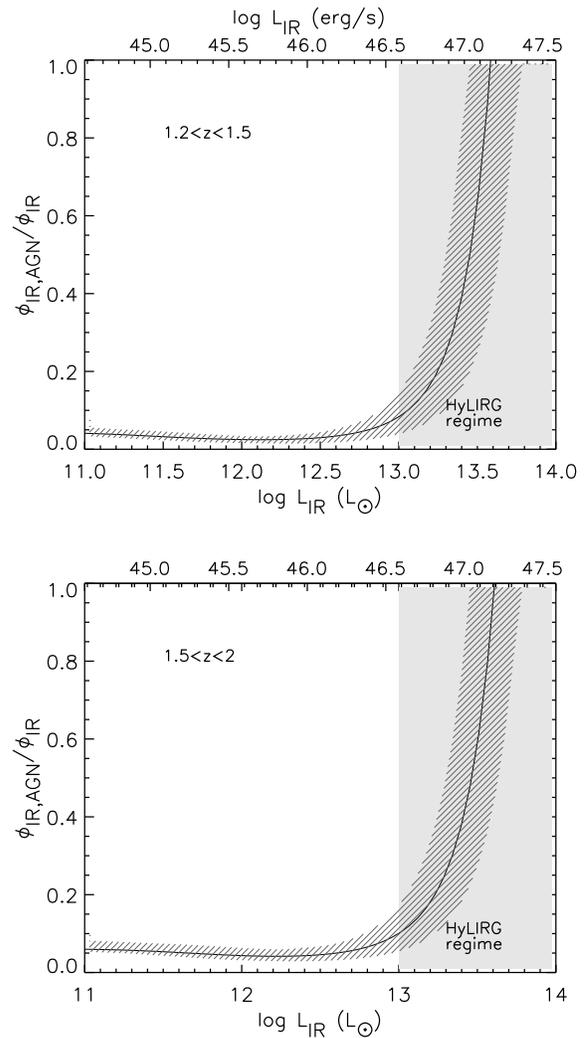,width=0.99\linewidth} 
\caption{Plotted here is the ratio of $\phi_{\rm IR, AGN}$ to $\phi_{\rm IR}$, which provides a simple estimate of the fraction of AGN-dominated sources as a function of $L_{\rm IR}$. The shaded outline to the curve represents the 1$\sigma$ error boundary calculated from the error boundaries of the luminosity functions in Fig. \ref{fig:LFs}. The 2 panels correspond to different redshift bins. The shaded vertical band in both panels is the HyLIRG regime ($L_{\rm IR}>10^{13}$\,L$_{\odot}$).}
\label{fig:AGN}
\end{figure}

\section{Method}
\label{sec:method}
Our investigation rests on the comparison of the infrared galaxy luminosity function and the AGN luminosity function. For the infrared galaxy luminosity function ($\phi_{\rm IR}$) we use the one presented in Gruppioni et al. (2013\nocite{Gruppioni13}; hereafter G13). $\phi_{\rm IR}$ is a function of $L_{\rm IR}$, which includes the total dust-reprocessed emission from stars and AGN. The uncertainties on $\phi_{\rm IR}$ from G13 are a combination of Poisson errors and photometric redshift uncertainties derived through Monte Carlo simulations. 

For the AGN luminosity function we use the absorption-corrected hard X-ray (2-10\,keV) AGN luminosity function from Aird et al. (2015\nocite{Aird15}; hereafter A15). This is significantly more complete than the optical AGN luminosity function (e.g. Richards et al. 2006\nocite{Richards06}; Ross et al. 2013\nocite{Ross13}) since it includes absorbed AGN. The X-ray AGN luminosity function is translated to an infrared AGN luminosity function ($\phi_{\rm IR, AGN}$) as follows: first, hard X-ray luminosity is converted to optical luminosity at 5100$\AA$ ($\nu L_{\nu, 5100}$), adopting the relation from Maiolino et al. (2007\nocite{Maiolino07}). Subsequently, to convert from $\nu L_{\nu, 5100}$ to infrared luminosity in the 8--1000$\mu$m range ($L_{\rm IR, AGN}$) we use the intrinsic AGN SED of Symeonidis et al. (2016; hereafter S16\nocite{Symeonidis16}). The errors on $\phi_{\rm IR, AGN}$ are Poisson, as are the original errors on the X-ray AGN luminosity function from A15. $L_{\rm IR, AGN}$ is the intrinsic IR luminosity of the AGN, i.e. it does not include the contribution of dust heated by starlight. $\phi_{\rm IR}$ and $\phi_{\rm IR, AGN}$ are monotonically decreasing functions of $L_{\rm IR}$ and $L_{\rm IR, AGN}$ respectively, over the luminosity range considered in this letter and  $\phi_{\rm IR} (L_{\rm IR}) \geq \phi_{\rm IR, AGN}(L_{\rm IR, AGN})$. 

\section{Results}
\label{sec:results}

The data and functional forms of $\phi_{\rm IR, AGN}$ and $\phi_{\rm IR}$ are shown in Fig. \ref{fig:LFs} in two redshift bins within the $1<z<2$ interval, corresponding to the peak in comoving infrared luminosity density. $z>1$ is where HyLIRGs first appear in the data of both luminosity functions, and we chose $z\sim$2 as a conservative upper limit relating to the reliability of photometric redshifts, because the bright end of the luminosity function is very sensitive to redshift mis-identification. $\phi_{\rm IR}$ is shown in the $1.2<z<1.7$, $1.7<z<2$ redshift bins; G13 fit the data with the function from Saunders et al. (1990), behaving as a power-law for $L<L_{\star}$ and as a Gaussian for $L>L_{\star}$ (see G13 for more details). $\phi_{\rm IR, AGN}$ is shown in the $1.2<z<1.5$, $1.5<z<2$ redshift bins and the data are fit with a double power-law model, whose parameters are themselves functions of redshift evaluated at the centre of the relevant bin (see A15 for more details); for the redshift bins of interest here, these are $z=1.35$ and $z=1.75$. As the G13 and A15 redshift bins are not precisely the same, we also evaluated the A15 LF model at the the centre of the G13 bins, namely $z=1.45$ and $z=1.85$. We found the mean shift in the A15 LF to be only about 0.1\,dex at the bright end, so we use the original redshift bins for $\phi_{\rm IR, AGN}$ in Fig. \ref{fig:LFs}, as the AGN luminosity densities were calculated in those bins in A15.

Fig. \ref{fig:LFs} shows that for $L_{IR}<10^{12}$~L$_{\odot}$, $\phi_{\rm IR}$ exceeds $\phi_{\rm IR, AGN}$ by more than 1\,dex. However, this difference decreases with increasing luminosity because $\phi_{\rm IR}$ declines faster than $\phi_{\rm IR, AGN}$ and at $L_{\rm IR} > 10^{13}$L$_{\odot}$ the two luminosity functions converge, with the measurements of $\phi_{\rm IR}$ and $\phi_{\rm IR, AGN}$ consistent at 1~$\sigma$. The models of $\phi_{\rm IR}$ and $\phi_{\rm IR, AGN}$ cross at log\, ($L_{\rm IR}$/L$_{\odot}$) $\sim$13.6 in both redshift ranges. 
 
The ratio of $\phi_{\rm IR, AGN}$ to $\phi_{\rm IR}$ provides a simple estimate of the fraction of AGN-dominated sources as a function of $L_{\rm IR}$. This ratio, as derived from the parametric model luminosity functions, is shown in Fig. \ref{fig:AGN}. In contrast to Fig. \ref{fig:LFs}, here we use the functional form of the A15 LF at the the centre of the G13 redshift bins. Fig. \ref{fig:AGN} shows that the contribution of the AGN to the total infrared luminosity and hence the fraction of AGN-dominated sources is small ($<$10 per cent) until the HyLIRG regime, where it undergoes a rapid increase and at log\,($L_{\rm IR}/\rm L_{\odot})>13.5$ the galaxy population becomes AGN dominated. 

\section{Discussion}
\label{sec:discussion}

 \begin{figure}
\epsfig{file=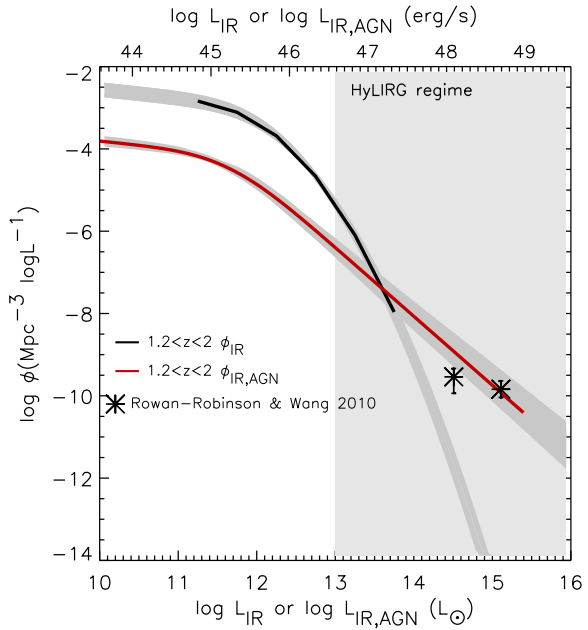,width=0.99\linewidth} 
\caption{The infrared galaxy luminosity function (black curve) and infrared AGN luminosity function (red curve) in the $1.2<z<2$ redshift bin (the redshift bins from Fig. \ref{fig:LFs} are combined by averaging the luminosity functions). The asterisks show the \textit{IRAS} HyLIRG luminosity function from Rowan-Robinson $\&$ Wang (2010).}
\label{fig:LFs_RRW10}
\end{figure}

\subsection{Implications for the infrared luminosity function}

In this work, we have used the AGN and galaxy infrared luminosity functions to explore what powers HyLIRGs at the peak of the cosmic star-formation history at $1<z<2$. We believe that this is the first time that the AGN luminosity function has been expressed as a function of \textit{intrinsic} infrared AGN power, enabling an estimate of the fraction of AGN dominated sources per luminosity bin. Our results show that the galaxy and AGN luminosity functions are initially offset in the $10^{11} - 10^{13}$\,L$_{\odot}$ luminosity range, but converge in the HyLIRG regime with the model luminosity functions crossing at about $10^{13.6}$\,L$_{\odot}$. By definition (see Section \ref{sec:method}) $\phi_{\rm IR}$ cannot be less than $\phi_{\rm IR, AGN}$, implying that beyond the crossover ($>10^{13.6}$\,L$_{\odot}$), the shape of one or both of the luminosity functions must be different to that predicted by the parametric models. 

The relatively small areas covered by the deep infrared surveys with suitable redshift information are a hindrance to adequately sampling the bright tail of the infrared luminosity function; Fig. \ref{fig:LFs} shows that the only HyLIRG data point in $\phi_{\rm IR}$ from G13 is at log\,$L_{\rm IR}/ \rm L_{\odot} <$13.5. On the other hand, the shape of the AGN luminosity function is well-established as a double power-law at all wavelengths (e.g. Dunlop $\&$ Peacock 1990\nocite{DP90}, Boyle, Shanks \& Peterson 1988\nocite{BSP88}, Page et al. 1997\nocite{Page97}) and the availability of large surveys such as the 2dF (Lewis, Glazebrook $\&$ Taylor 1998\nocite{LGT98}) and SDSS (York et al. 2000\nocite{York00}) have enabled precise measurements of the bright tail of the AGN luminosity function, e.g. Boyle et al. (2000)\nocite{Boyle00}, Ross et al. (2013)\nocite{Ross13}. 
The most natural resolution of the apparent crossing of $\phi_{\rm IR}$ and $\phi_{\rm IR, AGN}$, would therefore be for $\phi_{\rm IR}$ to flatten above the highest luminosities probed by G13 to match the shape of $\phi_{\rm IR, AGN}$. Thus the IR emission from log\,($L_{\rm IR}$/L$_{\odot}$)$>$13.5 galaxies is dominated by emission from AGN, and the luminosity function of galaxies in this regime is shaped by AGN rather than star-forming galaxies. 

We are able to test the prediction that $\phi_{\rm IR}$ must flatten at $L_{\rm IR} > 10^{13.5}$ with the $1<z<2$ \textit{IRAS} HyLIRG luminosity function presented in Rowan-Robinson $\&$ Wang (2010\nocite{RRW10}; hereafter RRW10). This is shown in Fig \ref{fig:LFs_RRW10}, where we combine the two redshift bins for $\phi_{\rm IR}$ and $\phi_{\rm IR, AGN}$ by averaging the luminosity functions, to give us $\phi_{\rm IR}$ and $\phi_{\rm IR, AGN}$ at $1.2<z<2$. To compare with the RRW10 results, we convert the abscissa of the RRW10 luminosity function from 60$\mu$m to total infrared luminosity using the S16 intrinsic AGN SED, a reasonable assumption given the SED shapes of HyLIRGs presented in RRW10, and that the AGN incidence at these luminosities is 100 per cent.
Fig \ref{fig:LFs_RRW10} shows that the space density of $1<z<2$ \textit{IRAS} HyLIRGs does indeed appear to be consistent with the space density of the extrapolated $\phi_{\rm IR, AGN}$. RRW10 themselves noted the apparent flattening of the luminosity function implied by their data at the highest luminosities, but were unable to identify the mechanism responsible. Here it is explained naturally by the transition to an entirely AGN-dominated population.

\subsection{On the nature of HyLIRGs}

Our study of the HyLIRG luminosity function indicates that in the
HyLIRG regime, the dominant power source in the galaxy population transitions 
from star-formation to AGN: figs. \ref{fig:LFs} and \ref{fig:AGN} show
that the total infrared emission of the most luminous HyLIRGs can be
accounted for entirely by AGN. An implication of this result, and a
condition necessary for it to hold, is that at $L_{\rm IR}>10^{13.5} \rm
L_{\odot}$ almost all HyLIRGs must host AGN. Observations of HyLIRGs
readily satisfy this condition: the incidence of AGN in infrared
galaxies rises as a function of infrared luminosity (e.g. Goto et
al. 2005\nocite{Goto05}; Kartaltepe et al. 2010\nocite{Kartaltepe10};
Yuan et al. 2010\nocite{Yuan10}; Goto et al. 2011a\nocite{Goto11a}), plateauing at 100 per cent in the HyLIRG regime. Indeed, as far as can be
ascertained, all known HyLIRGs host AGN (e.g. Rowan-Robinson
2000\nocite{RR00}; Farrah et al. 2002a\nocite{Farrah02a};
2002b\nocite{Farrah02b}; Ruiz et al. 2010\nocite{Ruiz10};
2013\nocite{Ruiz13}; Calder{\'o}n et al. 2016\nocite{Calderon16}; Toba
$\&$ Nagao 2016\nocite{TN16}).

Although the number of HyLIRGs studied in detail so far is small, it seems that the transition from star-formation to AGN power within the HyLIRG regime is consistent with the variation of HyLIRG spectral properties with luminosity. The most luminous HyLIRGs show, on average, more prominent AGN signatures: optical and X-ray spectra characteristic of AGN and SEDs that are AGN-dominated from the optical to (at least) the mid-IR (e.g. Ruiz, Carrera $\&$ Panessa 2007\nocite{RCP07}; Ruiz et al. 2010\nocite{Ruiz10}). 

We note that although the transition from SF to AGN dominated sources occurs in the HyLIRG regime at $z=1-2$, this could be different at other redshifts as the knee of the luminosity function evolves with redshift. It is plausible that at lower redshift the transition takes place at lower luminosities --- for example Nardini et al. (2010\nocite{Nardini}) suggest that, for $z<0.35$ ULIRGs with log\,$[L_{\rm IR}/\rm L_{\odot}]>12.7$, the fraction of AGN-dominated sources is about 40 per cent.

\subsection{Implications for galaxy star formation rates}
\label{sec:sfrs}
We have found that AGN can account for a significant part of the infrared emission in $13< \rm log (L_{\rm IR}/ L_{\odot})<13.5$ sources and the entire infrared emission at $L_{\rm IR}>10^{13.5}\, \rm L_{\odot}$. Assuming that the total infrared luminosity is linked to star-formation, $L_{\rm IR}=10^{13}\, \rm L_{\odot}$ and $L_{\rm IR}=10^{13.5}\, \rm L_{\odot}$ would translate to an SFR of 1700\,M$_{\odot}$/yr and 5400\,M$_{\odot}$/yr respectively, using the Kennicutt (1998\nocite{Kennicutt98}) SFR-$L_{\rm IR}$ relation. Our results show that, at least at $1<z<2$, SFRs claimed to be greater than 5400\,M$_{\odot}$/yr (e.g. Rowan-Robinson $\&$ Wang 2010\nocite{RRW10}; Rowan-Robinson et al. 2016\nocite{RR16}; Banerji et al. 2017\nocite{Banerji17}; Stacey et al. 2018\nocite{Stacey18}) should be scrutinised carefully because such objects are likely to be entirely powered by AGN. The dominance of AGN at those luminosities does not imply that SFRs$>$5400\,M$_{\odot}$/yr are implausible, rather than the existence of such SFRs cannot be established from the mere existence of galaxies in this luminosity range. 

Even sources with lower reported SFRs but still in the HyLIRG regime are likely to require a large correction due to the AGN contribution at infrared wavelengths. Although most authors use multi-component SED fitting to estimate HyLIRG star-formation rates (e.g. Rowan-Robinson et al. 2000\nocite{RR00}; Farrah et al. 2002b\nocite{Farrah02b}; Dai et al. 2012\nocite{Dai12}; Leipski et al. 2013\nocite{Leipski13}; 2014\nocite{Leipski14}; Fan et al. 2016\nocite{Fan16}; Duras et al. 2017\nocite{Duras17}), the far-IR has always been attributed to dust heated by star-formation. In Symeonidis et al. (2016) and Symeonidis (2017\nocite{Symeonidis17}) however, we showed that for galaxies hosting powerful AGN, the far-IR emission is likely to originate from kpc-scale dust heated by the AGN. 

We believe that it is imperative to consider diverse star-formation rate indicators for HyLIRGs, not associated with broadband photometry. Some studies which have examined gas masses, mid-IR spectral features, or extended soft X-ray emission in such sources (e.g. Yun $\&$ Scoville 1998\nocite{YS98}; Nandra $\&$ Iwasawa 2007\nocite{NI07}), have concluded that the implied SFRs are not sufficient to power the infrared emission. Our proposal that the most luminous infrared galaxies (at least at $1<z<2$) have lower SFRs than often suggested, could be an important step towards bridging the discrepancies between models and observations. High SFRs of the order of a few thousand M$_{\odot}$/yr or more have been difficult to explain; for example they are much higher than the estimated Eddington limited SFR (in the optically thick limit) of $\sim$1400\,M$_{\odot}$/yr for typical high redshift ULIRGs (e.g. Murray, Quataert $\&$ Thompson 2005\nocite{MQT05}). Moreover, it has been challenging for the $\Lambda$ Cold Dark Matter ($\Lambda$CDM) galaxy formation models to reproduce the most luminous dusty galaxies at high redshift without evoking a flat or top heavy initial mass function (IMF; e.g. Baugh et al. 2005\nocite{Baugh05}, Swinbank et al. 2008\nocite{Swinbank08}; Dav\'e et al. 2010\nocite{Dave10}).

\subsection{Implications for galaxy surveys}
Since the discovery of HyLIRGs from the \textit{IRAS} survey, the increased sensitivity of the \textit{WISE} survey has facilitated the discovery of an unprecedented number of hyperluminous sources  (e.g. Secrest et al. 2015\nocite{Secrest15}; Toba $\&$ Nagao 2016\nocite{TN16}). Here we have shown that the most luminous sources discovered by \textit{WISE}, or any other all-sky survey (at least at $1<z<2$), are likely to be AGN-powered. Since AGN completely dominate the emission in $L_{\rm IR}>10^{13.5}\,\rm L_{\odot}$ galaxies, a sample of HyLIRGs in this luminosity range may be perfect for finding out the obscured fraction of AGN at the highest luminosities and discriminating between the various torus models that have been put forward (e.g. Lawrence 1991\nocite{Lawrence91}; {H{\"o}nig} $\&$ T. Beckert 2007\nocite{HB07}; Lawrence $\&$ Elvis 2010\nocite{LE10}; Mateos et al. 2017\nocite{Mateos17}). Moreover, identifying the most luminous IR galaxies is a means of discovering AGN even if these are Compton thick, a significant step towards quantifying the contribution of the most heavily obscured AGN to the X-ray background (e.g. Akylas et al. 2012\nocite{Akylas12}; Buchner et al. 2015\nocite{Buchner15}).

\subsection{The robustness of our computed fraction of AGN-dominated sources} 

\subsubsection{The contribution of compton thick AGN}
The A15 luminosity function includes a component from Compton thick (CT) AGN, which affects its normalisation. Although the CT AGN fraction is uncertain, the value quoted in A15 is 34 per cent of the absorbed AGN, lower than other CT fractions quoted in the literature which range from 35 to 50 per cent of the whole AGN population. For example, Akylas et al. (2016\nocite{Akylas16}) argue that for a CT fraction that evolves as a function of redshift, a value as high as 50 per cent at $z\sim1$ would be consistent with the 2-10\,keV number counts. An evolving CT fraction is in agreement with the results of Brightman $\&$ Ueda (2012\nocite{BU12}), who find that the CT fraction increases with redshift, measuring 42$\pm7$ per cent per cent at $z\sim2$ (see also Vignali et al. 2014\nocite{Vignali14}). To find how our results would change, we used the upper end of the quoted CT fraction range, i.e. 50 per cent, to re-compute the fraction of AGN-dominated sources as a function of $L_{\rm IR}$. We find that the point at which the galaxy population becomes AGN-dominated shifts to log\,[$L_{\rm IR}/L_{\odot}] > 13.4$, 0.1\,dex lower than the value quoted in section \ref{sec:results}, but within the 1$\sigma$ confidence interval outlined in Fig \ref{fig:AGN}.

\subsubsection{The effect of SED choice}
The S16 intrinsic AGN SED we used to convert the X-ray AGN luminosity function to an IR AGN luminosity function (see section \ref{sec:method}) is arguably the most appropriate choice because it includes emission from AGN-heated dust in the torus as well as further afield, at kpc scales. S16 present a detailed comparison between the S16 SED and other AGN SEDs, finding that they all fall short of the S16 SED in the far-IR, be it because their derivation inherently assumes that AGN do not heat dust at kpc scales, or because their formulation only treats dust emission from the AGN torus. Nevertheless, we examine the effect of SED choice on our results, by re-computing the IR AGN luminosity function using the high luminosity AGN SED from Mullaney et al. (2011\nocite{Mullaney11}), which has the least far-IR emission out of the three SEDs they present and is similar to the Netzer et al. (2007\nocite{Netzer07}) AGN SED (see S16). We find that the AGN luminosity function is shifted leftward by 0.1\,dex which translates to a rightward shift in the curve of the fraction of AGN-dominated sources by about 0.07\,dex, within the 1$\sigma$ confidence interval shown in Fig \ref{fig:AGN}.

\section{Conclusions}
\label{sec:conclusions}

We have investigated what powers HyLIRGs at z$\sim$1-2, by examining the behaviour of the infrared AGN luminosity function in relation to the infrared galaxy luminosity function. The former corresponds to emission from AGN-heated dust only, whereas the latter includes emission from dust heated by AGN and starlight.
At the peak of cosmic SFR history, $z\sim 1-2$:
\begin{itemize}
\item The AGN and galaxy luminosity functions converge in the HyLIRG regime, where the measured space densities of galaxies and AGN become comparable. This implies a transition between galaxies primarily powered by star-formation to galaxies primarily powered by AGN, the transition locus being at log\,[$L_{\rm IR}/\rm L_{\odot}] \sim$13.5
\item The shape of the infrared galaxy luminosity function at the most extreme luminosities flattens to match that of the AGN luminosity function, implying that at the bright end the infrared galaxy luminosity function is shaped by AGN, not by star formation. 
\item HyLIRG SFRs calculated using broadband photometry are likely overestimated and require a correction due to the AGN contribution. In particular, galaxies which have infrared luminosities that would translate to SFRs of more than a few thousand M$_{\odot}$/yr are instead likely to be primarily powered by the AGN.

\end{itemize}

\bibliographystyle{mn2e}
\bibliography{references}

\end{document}